\documentclass[12pt,a4paper]{article}
\usepackage[cp866]{inputenc}
\begin{document}

\begin{center}
{\bf A.F. Krutov}\footnote{Samara State University.
E-mail: krutov@ssu samara.ru},
{\bf V.E. Troitsky}\footnote{D.V. Skobeltsyn Institute of Nuclear Physics,
Moscow State University.\\
E-mail: troitsky@theory.sinp.msu.ru}

{\large \bf FORM FACTORS OF COMPOSITE SYSTEMS \\
BY GENERALIZED WIGNER-ECKART THEOREM FOR POINCAR\'E  GROUP}

\end{center}

The relativistic approach to electroweak properties of
two-particle composite systems developed previously is
generalized here to the case of nonzero spin. This approach is
based on the instant form of relativistic Hamiltonian dynamics. A
special mathematical technique is used for the parametrization of
matrix elements of electroweak current operators in terms of form
factors. The parametrization is a realization of the generalized
Wigner--Eckart theorem on the Poincar\'e group, form factors are
corresponding reduced matrix elements and they have the sense of
distributions (generalized functions). The electroweak current
matrix element satisfies the relativistic covariance conditions
and in the case of electromagnetic current it also automatically
satisfies the conservation law.
\vspace{10mm}


\newpage

\section{Introduction}

The development of correct quantitative methods of calculation
of composite--particle structure is an important line of
investigations in particle physics.
The modern experimental researches of the particle structure are performed
with such an accuracy and for such energies that it calls for the
relativistic theoretical method of interpretation of results.
The important feature of these methods is the extraction of kinematical
parts and invariant parts (form factors) from the current matrix elements.
A representation of the matrix element in terms of form factors is called
the parametrization of the current operator matrix element.
Let us remark that it is just form factors -- Lorentz invariant
functions -- that are extracted from scattering data usually.

In the relativistic theory of the description of composite
systems, it is possible to identify two main but very
different approaches (see e.g. \cite{BaK95}
and the pictorial scheme on Fig. 11 in the review \cite{GiG02}).

The first is the method of field theory. Based on the principles
of quantum field theory -- quantum chromodynamics (QCD) -- it is
rightly regarded as the most consistent approach to the solution
of this problem. However, standard perturbative QCD  gives
sufficiently reliable computational prescriptions only for the
description of so--called "hard" processes, which are
characterized by large momentum transfers, and it does not
permit the calculation of characteristics determined by "soft"
processes. Moreover, there are strong indications
that perturbative QCD is not valid for the
description of the currently existing experimental facts in
exclusive processes.  This applies, in particular, to the
description of the elastic form factors of such well--studied
composite systems as pion, nucleon and deuteron.

The second method in the relativistic theory of composite
systems that is gaining ac\-cep\-tan\-ce  in recent years is based on
the direct realization of the algebra of the Poincar\'e group on
the set of dynamical observables on the Hilbert state space of
the system. This approach is called the theory of direct
interaction, or relativistic Hamiltonian dynamics (RHD). The discussion of this
approach can be
found e.g. in \cite{GiG02,KeP91}.

The method is based on the quite different starting positions than the
first one. It is one of the reasons why the results of these approaches
are difficult to compare.
We shall not focus ourselves on the details of RHD but we refer reader
to the paper \cite{KrT02} and references therein.

Although the above-mentioned methods of the description of composite systems are
different
enough, it is necessary to solve in both cases three highly important
problems.

1) It is necessary to give the relativistic recipe for taking into
account the nonperturbative effects. Notice that in the case of RHD, in
the framework of which we shall operate, these
effects are contained in the structure of method directly
\cite{BaK95,KrT02}.

2) It is necessary to construct the matrix element of electroweak current with
correct
transformation properties and to perform the procedure of the relativistic
parametrization of
matrix elements that is to extract the relativistic invariant form factors.

3) It is necessary to take into account the fact that the form factors
of composite systems have, generally speaking, the sense of
distributions(generalized functions), even in the nonrelativistic case.

In simple cases the parametrization can be obtained with reasonable facility
from general
considerations. However, in more complicated cases, for example, in
the case of composite systems with arbitrary total angular momentum one needs a
special
mathematical method.

The aim of the present paper is the construction of a general
parametrization method  for the case of composite systems with nonzero
values of total spin.

The method was announced in our brief report \cite{KrT03prc}. Now we shall give
the consistent
presentation of our method which makes it possible to apply it to the
specific composite systems. Our method is similar in some sense to the
method of canonical parametrization of one-particle matrix elements of
field operators \cite{ChS63}. It should be particularly emphasized that we
are dealing now with relativistic Hamiltonian dynamics rather than with
field theory as in Ref. \cite{ChS63}.

Conceptually, our method of parametrization is a
realization of the Wigner--Eckart theorem for the Poincar\'e
group and form factors are the reduced matrix elements.
This theorem is used extensively in the different fields of theoretical and
mathematical
physics from well known theory of angular momentum to the Hopf algebras and
quantum superalgebras \cite{Moz04}.

In the present paper we use the term -- generalized Wigner--Eckart
theorem for the Poincar\'e group -- when form factors of composite systems
are considered as generalized functions. The necessity of such
consideration was shown in Refs. \cite{BaK95,KrT02} in the study of form
factors of composite systems with zero total spin.

The paper is organized as follows.

In Sec.II the canonical parametrization of local operator matrix
elements between one--particle states of arbitrary nonzero spin
is described. This parametrization presents the extraction of
the reduced matrix element on the Poincar\'e group. The
electromagnetic current matrix element is derived for the spin
1/2 particle.

In Sec.III we show how to construct the electromagnetic current
operator for composite system of two free particles of spin 1/2.
The current matrix element is obtained in the basis where the
center--of--mass motion is separated. The electromagnetic
properties of the system are defined by reduced matrix
elements, or the so called free two--particle form factors,
these form factors being generalized functions. This means
that, for example, the static properties of the system are given
by the weak limits as $Q^2\to$ 0
($Q^2 = -q^2\;,\;q$ is the momentum transfer).
A special attention is paid to the case of total spin
one and zero total orbital momentum. In
this case the electromagnetic properties are defined by four
free two--particle form factors -- charge, quadrupole,
magnetic and the magnetic quadrupole form factor of the second kind
(see the review ~\cite{DuT90} and the references therein).

In Sec.IV the procedure of the construction of the electromagnetic current
matrix element is
developed for the case of two interacting particles. Each step
of the calculation in the developed method
remains Lorentz covariant and
conserves the current. The composite system form factors are
derived as some integral representations in terms of the wave
functions obtained in the frame of RHD.

In Sec.V the results are summarized and the conclusion is given.

\section{Parametrization of one--particle operator
matrix elements}

Let us describe now a general method of canonical
parameterization of local operator matrix elements. In the present paper we are
dealing with the matrix elements diagonal in total angular
momentum only.

The main idea of the parametrization can be formulated as
follows. Using the variables entering the state vectors
which define the matrix elements one has to construct two
types of objects.

1. A set of linearly independent matrices which are
Lorentz scalars (scalars or pseudoscalars). This set describes
transition matrix elements non-diagonal in spin projections
in the initial and finite states, as well as the properties
defined by the discrete space--time transformations.

2. A set of linearly independent objects with the same
tensor dimension as the operator under consideration
(for example, four--vector, or four--tensor of some rank). This
set describes the matrix element behavior under the action of Lorentz group
transformations.

The operator matrix element is written as a sum of all possible
objects of the first type multiplied by all possible objects of the second
type. The coefficients in this representation as a sum are just the
reduced matrix elements -- form factors.

The obtained representation is then modified with the use of
additional conditions for the operator, such as the conservation
laws, for example. In order to satisfy these additional conditions in
some cases some of the coefficients (form factors) occur to be zero.

To demonstrate this let us consider the parameterization of the
matrix elements taken between the states of a free particle of
mass $M$ in different simple cases. Let us normalize the state
vectors as follows:
\begin{equation}
\langle\,\!\vec p\,,m\,|\,\vec p\,'\,,m'\,\!\rangle =2p_0\,\delta (\vec p - \vec
p\,')\,\delta
_{mm'}\;,
\label{normg}
\end{equation}
here $\vec p\;,\;\vec p\,'$ are three--momenta of particle,$p_0 = \sqrt{M^2 + \vec p\,^2}$,
$m\;,\;m'$ are spin
projections.

Let us consider first the parameterization of the matrix
element of scalar operator
$A(x)$ taken between the states of free particle with zero
spin. Because of translation invariance it is sufficient to
consider $A(0)$. Using the variables of state vectors in the
initial and final states one can construct two linearly
independent scalars:
$p^2 = {p'}\,^2 = M^2\;,\;(p - p')^2 = q^2 = - Q^2$.
Only the second one -- the momentum transfer
square -- is nontrivial. Ignoring the trivial dependence on $M$ we can write:
\begin{equation}
\langle\,\vec p,\,M|A(0)|\,\vec p\,',\,M\,\rangle = f(Q^2)\;.
\label{A(0)}
\end{equation}

Let us consider now the matrix element of a scalar operator
between the states of a particle with spin $j$:
\begin{equation}
\langle\,\vec p,\,M,\,j,\,m\,|A(0)|\,\vec p\,',\,M,\,j,\,m'\,\rangle\;.
\label{A(0)j}
\end{equation}
The tensor dimension of the operator is not changed,
nevertheless, the matrix element is now a matrix in spin
projections in the initial and finite states.

Let us construct the set of Lorentz scalars -- linearly
independent matrices in spin projections to be used for
the construction of the representation of Eq.~(\ref{A(0)j}).  Let
us use the covariant spin operator $\Gamma^\mu(p)$.
In the rest frame this operator coincides with the particle spin
operator:
\begin{equation}
\Gamma^0(0) = 0\;,\quad
\vec\Gamma(0) = M\,\vec j\;,\quad [\,j_i\,,j_k\,] =
i\,\varepsilon_{ikl}\,j_l\;.
\label{Lspin}\end{equation}
The covariant spin operator can be defined with the use of the
Pauli-Lubanski vector $w^\mu$ \cite{Nov72}. In terms of matrix
elements we have:
$$
\Gamma^\mu_{mm'}(p) = \langle\,\vec p\,,m|w^\mu|\vec p\,,m'\rangle=
\langle\,0\,,m|\hat U^{-1}(\Lambda_p)w^\mu \hat U(\Lambda_p)|0\,,m'\rangle =
$$
\begin{equation}=
\left(\Lambda_p\right)^\mu_\nu\langle\,0\,,m|w^\nu|0\,,m'\rangle
= \left(\Lambda_p\right)^\mu_\nu\,\Gamma^\nu_{mm'}(0)\;.
\label{Gw}
\end{equation}
Here $\Lambda_p$ is the boost for the transformation from
the rest frame of the particle to the laboratory frame,
$\hat U(\Lambda_p)$ is the corresponding representation
operator. The matrix
$\Lambda_p$ can be written explicitly using the matrix for
Lorentz transformation of the vector $p'$ into the vector $p$:
\begin{equation}
\Lambda ^\mu _{\;\nu} = \delta^\mu_\nu + \frac{2}{M^2}p^\mu\,p'_\nu -
\frac{(p^\mu + {p'}\,^\mu)(p_\nu + p'_\nu)}{M^2 + p^\lambda\,p'_\lambda}\;.
\label{Lambda}
\end{equation}
In our case $\Lambda_p$ is given by Eq.~(\ref{Lambda}) with
$p'=(M\;,\;0\;,\;0\;,\;0)$.

Using the explicit form of $\Lambda_p$ we obtain:
\begin{equation}
\Gamma_0(p) = (\vec p\vec j)\;,\quad
\vec \Gamma(p) =  M\,\vec j + \frac {\vec p(\vec p\vec j)}{p_0 + M}\;,
\quad \Gamma^2 = -M^2\,j(j+1)\;.\label{ Gamma mu}
\end{equation}

As is known \cite{Nov72}, spin transforms under the action of
Lorentz group following the so called little group which is
isomorphic to the rotation group, that is the corresponding
transformations are realized by the matrices of three
dimensional rotations.

So, the 4-spin operator is transformed under Lorentz
transformations $p^\mu = \Lambda^\mu_{\;\nu}\,p'\,^\nu$
in the following way:
\begin{equation}
\Gamma^\mu(p) = \Lambda ^\mu_{\;\nu}\,D^j(p,\,p')\,
\Gamma^\nu(p')\,D^j(p',\,p)\;.
\label{Lambda Gamma mu}
\end{equation}
Using Eq.~(\ref{Lambda Gamma mu})
one can show directly that the matrix elements of the operators
$D^j(p,\,p')\Gamma^\mu(p')$ and $\Gamma^\mu(p)D^j(p,\,p')$
transformed as 4--pseudovectors, the matrix elements of the
operators $D^j(p,\,p')p_\mu\Gamma^\mu(p')$ and
$p'_\mu\Gamma^\mu(p) D^j(p,\,p')$ -- as 4-pseudoscalars.

We shall construct the set of linearly independent Lorentz--scalar
matrices using the vectors
$p^\mu,\;{p'}^\mu$ and the pseudovector
$D^j(p,\,p')\Gamma^\mu(p')$.
Note, that the pseudovector
$\Gamma^\mu (p)D^j(p,\,p')$ does not enter the
decomposition,
being linearly dependent. One can show this fact using the
relation (\ref{Lambda Gamma mu}) and the explicit form
Eq.~(\ref{Lambda}) of $\Lambda ^\mu _{\;\nu}$ transforming  $p'$
into $p$.  It is easy to obtain:
\begin{equation}
\Gamma^\mu (p)D^j(p,\,p') = D^j(p,\,p')\left\{\Gamma^\mu (p')\right.
\left. - \frac {p^\mu +{p'}^\mu}{M^2 + p_\mu {p'}^\mu}\,
\left[p_\nu \Gamma^\nu (p')\right]\right\}\;.
\label{Gamma D = D Gamma}
\end{equation}
As ${p'}_\mu \Gamma^\mu (p') = 0$, the set in question of
linearly independent matrices in spin projections of
the initial and the final states giving the set of independent
Lorentz scalars is presented by $2j + 1$ quantities:
\begin{equation}
D^j(p,\,p')\,(p_\mu \Gamma^\mu (p'))^n\;,\quad n = 0,1,\ldots ,2j\;.
\label{pseud}
\end{equation}
The number of linearly independent scalars in
Eq.~(\ref{pseud}) is limited by the fact that the product of more
than $2j$ elements
$\Gamma^\mu (p')$ reduces as it is linearly dependent.
If $n$ is even, then the obtained quantities are scalars, if $n$
is odd then they are pseudoscalars. The current matrix element (\ref{A(0)j})
is represented by the linear combination of these linearly
independent Lorenrz scalars. The coefficients in this
combination $f_n(Q^2)$ are just form factors.
These form factors are invariant under rotations and so they do
not depend on spin projections. So, they depend upon only one
scalar combination of variables -- the momentum--transfer squared.

For self--adjoint operator $A(0)$ a minor modification of the set
(\ref{pseud}) is necessary: in the scalar product in
Eq.~(\ref{pseud}) the factor $i$ appears.
So, now the current matrix element
(\ref{A(0)j}) can be written in the form:
$$
\langle\,\vec p,\,M,\,j,\,m\,|A(0)|\,\vec p\,',\,M,\,j,\,m'\,\rangle =
$$
\begin{equation}
= \sum _{n=0}^{2j}\sum_{m''=-j}^{j}\,\langle\,m|D^j(p,\,p')|m''\rangle
\langle\,m''\,|\{ip_\mu\Gamma^\mu (p')\}^n|\,m'\,\rangle\,f_n(Q^2)
\label{DA(0)j= f_n}
\end{equation}
The obtained operator is self--adjoint. This can be easily
shown using the following relation:
\begin{equation}
p'_\mu \Gamma^\mu(p)D^j(p,\,p') = -D^j(p,\,p')p_\mu \Gamma^\mu (p')
\label{p Gamma D = - Dp Gamma}
\end{equation}
In the case of a scalar operator the values $n$ will be even and
for a pseudoscalar they will be odd.

Let us consider now the 4-vector operator $j_\mu(0)$.
To parametrize the matrix element one needs a set of
quantities of the appropriate tensor dimension. Using
the variables entering the particle state vectors one can
construct one pseudovector $\Gamma^\mu (p')$ and three
independent vectors:
\begin{equation}
K_\mu = (p - p')_\mu = q_\mu\,,\quad K'_\mu = (p + p')_\mu \,,\quad
R_\mu =\epsilon _{\mu \,\nu \,\lambda\,\rho}\, p^\nu \,p'\,^\lambda\,
\Gamma^\rho (p')\;.
\label{kk'RG}
\end{equation}
Here $\epsilon _{\mu \,\nu \,\lambda\,\rho}$ is a completely
anti-symmetric pseudo-tensor in four dimensional space-time with
$\epsilon _{0\,1\,2\,3}= -1$.

The operator matrix element contains the matrix elements of the
listed quantities multiplied by
$D^j(p,\,p')$ from the left. Each of such products is to be
multiplied by the sum of linearly independent scalars
constructed while obtaining the parameterization(\ref{DA(0)j= f_n}):
$$
\langle\,\vec p,\,M,\,j,\,m\,|j_\mu(0)|\,\vec p\,',\,M,\,j,\,m'\,\rangle =
$$
\begin{equation}
= \sum_{m''}\,\langle\,m|D^j(p,\,p')|m''\rangle
\langle\,m''|\,F_1\,K'_\mu + F_2\,\Gamma_\mu (p')
+ F_3\,R_\mu + F_4\,K_\mu |m'\rangle\;,
\label{<|j|>=F_is}
\end{equation}
where
\begin{equation}F_i = \sum _{n=0}^{2j}\,f_{in}(Q^2)(ip_\mu\Gamma^\mu(p'))^n\;.
\label{Fi}
\end{equation}

Let us impose some additional conditions on the operator.

1. Let us require the operator to be self--adjoint.

One can
check this condition by use of Eq.~(\ref{p Gamma D = - Dp Gamma}).  In the
right--hand side (r.h.s.) of
Eq.~(\ref{<|j|>=F_is}) we need to modify slightly the vector
multiplied by $F_2$.  The new vector is a linear combination of
the 4--vectors entering Eq.~(\ref{<|j|>=F_is}) and has the
following form:
\begin{equation}
\Gamma^\mu(p')\;\to\;
\Gamma^\mu(p') - \frac{{K'}\,^\mu}{{K'}^2}\,\left(p_\mu
\Gamma^\mu(p')\right)\;.
\label{G-K}
\end{equation}
The terms
containing $F_2$ and $F_3$ are modified in the following way:
\begin{equation}
F_i\,A^\mu\;\to\;\frac{1}{2}\left(F_i\,A^\mu + A^\mu\,F_i\right) =
\left\{F_i\,A^\mu\right\}_+\;,\quad i=2,3\;.
\label{+}
\end{equation}
Here $A^\mu$ are the vectors entering Eq.~(\ref{<|j|>=F_is}) and
changed following Eq.~(\ref{G-K}).  The terms containing $F_3$
and $F_4$ have to be multiplied by $i$.

2. It is useful to modify the parametrization so as
to make the vectors -- multipliers of
$F_i$ -- orthogonal to each other.

This results in the following form of the vector multiplied by
$F_2$ (\ref{G-K}):
\begin{equation}\Gamma^\mu(p') - \left(\frac{{K'}\,^\mu}{{K'}^2} +
\frac{{K}\,^\mu}{{K}^2}\right)\left(p_\mu \Gamma^\mu(p')\right)\;.
\label{G-KKs}
\end{equation}

3. Let us impose the condition of parity conservation.

This condition is satisfied if all the terms in the sum
(\ref{Fi}) contain an even number of pseudovector factors $\Gamma^\mu$.
So, the sums in $F_1$ and $F_4$ are over even
$n:\;2j\;\geq n\;\geq \;0$, in $F_3$ over even $n:\;2j - 1\;\geq
\;n\;\geq \;0$, and in $F_2$ over odd $n:\; 2j -1\;\geq\;n\;>\;0$.

4. Let us impose the conservation condition
$j_\mu K^\mu = j_\mu q^\mu= 0$.

It is easy to see that this condition is satisfied only if
$F_4 = 0$.

So, finally we have the following parametrization of the matrix element:
$$
\langle\,\vec p,\,M,\,j,\,m\,|j_\mu(0) |\,\vec p\,',\,M,\,j,\,m'\,\rangle
= \sum_{m''}\,\langle\,m|D^j(p,\,p')|m''\rangle\times
$$
\begin{equation}
\times
\langle\,m''|\,F_1K'_\mu + \left\{F_2\left[\Gamma_\mu (p')\right.\right.
\left.\left.
- (p_\mu \Gamma^\mu(p'))
\left(\frac {K'_\mu}{{K'}^2}+\frac {K_\mu}{{K}^2}\right )\right]\right\}_+ +
i\left\{F_3\,R_\mu\right\}_+|m'\rangle,\label{<|j|>=F_i}
\end{equation}
This construction can be used, for example, to obtain the
electromagnetic current matrix element in the case of particle
with spin 1/2.

Now let us list the conditions for the electromagnetic current operator to
be fulfilled in relativistic case (see Ref.~\cite{KrT02} and the
references therein).

1. Lorentz covariance:
\begin{equation}
\hat U^{-1}(\Lambda )\hat j^\mu (x)\hat U(\Lambda ) =
\Lambda ^\mu_{\>\nu}\hat j^\nu (\Lambda ^{-1}x)\;.
\label{UjmuU}
\end{equation}
Here $\Lambda $ is a Lorentz--transformation matrix,
$\hat U(\Lambda ) $ is an operator of the unitary
representation of the Lorentz group.\\

2. Invariance under translation:
\begin{equation}
\hat U^{-1}(a)\hat j^\mu(x)\hat U(a) = \hat j^\mu(x-a)\;.
\label{UajmuUa}\end{equation}
Here $\hat U(a)$  is an operator of unitary
representation of the translation group.\\

3. Current conservation law:
\begin{equation}
[\,\hat P_\nu\,\hat j^\nu(0)\,] = 0\;.
\label{Pj=0}
\end{equation}
In terms of matrix elements
$\langle\,\hat j^\mu\,\rangle$
the conservation law can be written in the following form:
\begin{equation}q_\mu\,\langle\,\hat j^\mu(0)\,\rangle  = 0\;.
\label{Qj=0}
\end{equation}
Here $q_\mu$ is four-vector of the momentum transfer.\\

4. Current--operator transformations under space--time
reflections:
$$
\hat U_P\left(\,\hat j^0(x^0\,,\vec x)\,,\hat{\vec j}(x^0\,,\vec x)\right)
\hat U^{-1}_P
=
\left(\,\hat j^0(x^0\,,-\,\vec x)\,,-\,\hat{\vec j}(x^0\,,-\,\vec x)\right)\;,
$$
\begin{equation}
\hat U_R\,\hat j^\mu(x)\,\hat U^{-1}_R = \hat j^\mu(-\,x)\;.
\label{UpUr}
\end{equation}
In Eq.~(\ref{UpUr}) $\hat U_P$ is the unitary operator for the
representation of space reflections and
$\hat U_R$ is the anti-unitary operator of the representation of
space-time reflections $R = P\,T$.

Our parametrization satisfies all the
conditions listed above.
The parametrization (\ref{<|j|>=F_i})
for the electromagnetic current in the case of particle with
spin 1/2 has the following form:
$$
\langle\,\vec p,\,M,\,\frac{1}{2},\,m\,|\,j_\mu(0)\, |\,\vec
p\,',\,M,\,\frac{1}{2},\,m'\,\rangle =
$$
\begin{equation}
= \sum_{m''}\,\langle\,m|D^{1/2}(p,\,p')|m''\rangle
\langle\,m''|\,f_{10}(Q^2)\,K'_\mu +
if_{30}(Q^2)\,R_\mu|m'\rangle\;,
\label{<|j|>=K+R}
\end{equation}
The form factors $f_{10}(Q^2)= f_1(Q^2)$ and $f_{30}(Q^2)=f_2(Q^2)$  are the
electric and the magnetic form factors of the particle, respectively.
In analogous way (although a little more cumbersome) one can
obtain the matrix elements of the operators of higher
tensor dimension.


\section{Parametrization of matrix elements of
two--particle\\
 electromagnetic current operator}
To describe the properties of the system of interacting
constituents in our approach it is necessary to have the reduced
matrix elements on Poincar\'e group (the form factors) which
describe the properties of the composite system of free
constituents. In this section we generalize the method of
parametrization of the previous section to the case of such free
systems.

Let us consider a system of two free particles with spins 1/2 and let us
parametrize the matrix element describing the transitions in
this system. Let us construct, for example, the electromagnetic
current operator matrix element. The matrix element can be taken
between the following two--particle state vectors:
\begin{equation}
|\,\vec p_1\,,m_1;\,\vec p_2\,,m_2\,\!\rangle =
|\,\vec p_1\,,m_1\,\!\rangle\otimes |\,\vec p_2\,,m_2\,\rangle\;.
\label{p1m1p2m2}
\end{equation}
Here $\vec p_1\;,\;\vec p_2$ are three--momenta of particles,
$m_1\;,\;m_2$ are spin projections on the $z$ axis.
The one--particle state vectors are normalized by
Eq.~(\ref{normg}).

As well as the basis (\ref{p1m1p2m2}) one can choose another
set of two-particle state vectors
where the motion of the
two-particle center of mass is separated:
$$
|\,\vec P,\;\sqrt {s},\;J,\;l,\;S,\;m_J\,\rangle\;,
$$
$$
\langle\,\vec P,\;\sqrt {s},\;J,\;l,\;S,\;m_J
|\,\vec P\,',\;\sqrt {s'},\;J',\;l',\;S',\;m_{J'}\,\rangle =
$$
\begin{equation}
= N_{CG}\,\delta^{(3)}(\vec P - \vec P\,')\delta(
\sqrt{s} - \sqrt{s'})\delta_{JJ'}\delta_{ll'}\delta_{SS'}\delta_{m_Jm_{J'}}\;,
\label{bas-cm}
\end{equation}
$$
N_{CG} = \frac{(2P_0)^2}{8\,k\,\sqrt{s}}\;,\quad
k = \frac{\sqrt{\lambda(s\,,\,M_1^2\,,\,M_2^2)}}{2\,\sqrt{s}}\;,
$$
here $P_\mu = (p_1 +p_2)_\mu$, $P^2_\mu = s$, $\sqrt {s}$
is the invariant mass of the two-particle system,
$l$ is the orbital angular momentum in the center--of--mass
frame (c.m.), $S$ is the total spin in the c.m., $J$ is the
total angular momentum with the projection $m_J$; $M_1\;,\;M_2 $
are the constituent masses, and
$\lambda (a,b,c) = a^2 + b^2 + c^2 - 2(ab + bc + ac)$.

The basis (\ref{bas-cm}) is connected with the basis
(\ref{p1m1p2m2})
through the Clebsh--Gordan decomposition
for the Poincar\'e group. Here we write the decomposition in
a little more general form than in
Ref.~\cite{KrT02}:
$$
|\,\vec P,\;\sqrt {s},\;J,\;l,\;S,\;m_J\,\rangle =
$$
\begin{equation}
= \sum_{m_1\;m_2}\,\int \,\frac {d\vec p_1}{2p_{10}}\,
\frac {d\vec p_2}{2p_{20}}\,|\,\vec p_1\,,m_1;\,\vec p_2\,,m_2\,\rangle
\langle\,\vec p_1\,,m_1;\,\vec p_2\,,m_2\,|
\,\vec P,\;\sqrt {s},\;J,\;l,\;S,\;m_J\,\rangle\;,
\label{Klebsh}
\end{equation}
where
$$
\langle\,\vec p_1\,,m_1;\,\vec p_2\,,m_2\,|\,\vec P,\;\sqrt {s},
\;J,\;l,\;S,\;m_J\,\rangle
= \sqrt {2s}[\lambda
(s,\,M_1^2,\,M_2^2)]^{-1/2}\,2P_0\,\delta (P - p_1 - p_2)\times
$$
$$
\times \sum
\langle\,m_1|\,D^{1/2}(p_1\,,P)\,|\tilde m_1\,\rangle
\langle\,m_2|\,D^{1/2}(p_2\,,P)\,|\tilde m_2\,\rangle\times
$$
$$
\times
\langle\frac{1}{2}\,\frac{1}{2}\,\tilde m_1\,\tilde
m_2\,|S\,m_S\,\rangle\,Y_{lm_l}(\vartheta\,,\varphi )\,
\langle S\,l\,m_s\,m_l\,|Jm_J\rangle\;.
$$
Here $\vartheta\;,\;\varphi$ are the spherical angles of the
vector $\vec p = (\vec p_1 - \vec p_2)/2$
in the c.m., $Y_{lm_l}(\vartheta,\varphi)$ are
spherical harmonics,
$\langle1/2\,1/2\,\tilde m_1\,\tilde m_2\,|S\,m_S\,\rangle\;,\;
\langle S\,l\,m_s\,m_l\,|Jm_J\rangle$ are Clebsh--Gordan
coefficients for the group $SU(2)$, $D^j$ are
the known rotation matrices to be used for correct relativistic
invariant spin addition, the summation is over
${\tilde m_1,\tilde m_2,m_l,m_S}$.

The decomposition in spherical harmonics and angular momenta summation
in Eq.~(\ref{Klebsh}) are performed in the c.m. and
the result is shifted to an arbitrary frame by use of
$D$--functions.

The electromagnetic current matrix element for the system of two
free particles taken in the basis
(\ref{p1m1p2m2}) can be written as a sum of the one--particle
current matrix elements:
$$
\langle\vec p_1,m_1;\vec p_2,m_2|j_\mu^{(0)}(0)|
\vec p\,'_1,m'_1;\vec p\,'_2,m'_2\rangle =
$$
\begin{equation}
= \langle\vec p_2,m_2|\vec p\,'_2,m'_2\rangle \langle\vec p_1,m_1|j_{1\mu}(0)|
\vec p\,'_1,m'_1\rangle + (1\leftrightarrow 2)\;.
\label{j=j1+j2}
\end{equation}
Each of the one--particle current matrix elements in
Eq.~(\ref{j=j1+j2}) can be written in terms of form factors as in
Section II. In the case of the particles with spin 1/2 we make
use of Eq.(\ref{<|j|>=K+R}). So, in this case the
electromagnetic properties  of the system are defined by the
form factors $f_{1}\;,\;f_{2}$.

Now let us construct the electromagnetic current matrix element
for the system of two free particles in the basis
(\ref{bas-cm}) following the previous Section. Let us consider
first a simple case
$J=J'=S=S'=l=l'=0$. (We omit these variables in the state
vectors.) This set of quantum numbers appears, for
example, in the case of pion \cite{KrT02}. Now there is no pseudovector
$\Gamma^\mu$, but along with the scalar $(P - P')^2 = -\,Q^2$
two additional nontrivial scalars do appear $s' = P'\,^2$ and
$s = P^2$ -- the invariant mass squares for the free two--particle
system in the initial and in the final states.  So, the form
factors entering the parametrization are functions of the
variables $Q^2\;,\;s\;,\;s'$. The current matrix element is
presented by the linear combination of the four--vectors $P_\mu$
and $P'_\mu$:
\begin{equation}
\langle\vec P,\sqrt s\mid j_\mu^{(0)}(0) \mid\vec P',\sqrt{s'}\rangle
=  (P_\mu + P'_\mu)\,g_1 (s,Q^2,s') + (P_\mu - P'_\mu)\,g_2 (s,Q^2,s')\;.
\label{<|j|>=Pg1Pg2}
\end{equation}

Making use of the conservation condition (\ref{Qj=0})
\begin{equation}
j_\mu^{(0)}(0)(P - P')^\mu = 0\;,
\label{conserv}
\end{equation}
we can write the parametrization in the form:
\begin{equation}
\langle\vec P,\sqrt s\mid j_\mu^{(0)}(0)\mid \vec P',\sqrt{s'}\rangle
=  A_\mu (s,Q^2,s')\;g_0 (s,Q^2,s')\;.
\label{<|j|>=A mu g0}
\end{equation}
Here $g_0 (s,Q^2,s')$ is the reduced matrix element. We will
refer to this invariant as to the free two--particle form
factor. The vector
$A_\mu (s,Q^2,s')$ is defined by the current transformation
properties (the Lorentz covariance and the conservation law):
\begin{equation}
A_\mu (s,Q^2,s')
= (1/Q^2)[(s-s'+Q^2)P_\mu + (s'-s+Q^2) P\,'_\mu]\;.
\label{Amu}
\end{equation}

The free two--particle form factor can be expressed in terms of
the one--particle form factors (\ref{<|j|>=K+R}).
To do this one has to perform in
Eq.~(\ref{<|j|>=A mu g0}) the Clebsh--Gordan decomposition of the
irreducible representation (\ref{bas-cm}) into the direct
product of two irreducible representations (\ref{p1m1p2m2}),
(\ref{Klebsh}) and to take into account
Eqs.~(\ref{<|j|>=K+R}), (\ref{j=j1+j2}).  As the form factors are
invariants one can perform the integration in Eq.~(\ref{<|j|>=A
mu g0}) in the coordinate frame with $\vec P\,' = 0\,$,
$\vec P = (0,\,0,\,P)$.  The explicit form of $g_0 (s,Q^2,s')$ in the
case of two particles with spin 1/2 and mass $M$ can be found in
Ref.~\cite{KrT02}.

Let us perform the analogous parameterization for the set of
quantum numbers in the basis
(\ref{bas-cm}) in the case $J\;,\;J'\ne$ 0.
In the following, we will take $J = J'$.

The Lorentz covariant properties of the matrix element are
defined (in analogy with
Eq.~(\ref{kk'RG})) by three 4--vectors and one pseudovector:
\begin{equation}
K'_\mu = (P + P')_\mu\;,\quad K_\mu = (P - P')_\mu \;,
R_\mu = \epsilon _{\mu \nu \lambda \rho}P^\nu P'^\lambda
\Gamma^\rho (P')\;, \quad \Gamma_\mu(P')\;.
\label{K'KRG}
\end{equation}
The pseudovector $\Gamma_\mu(P)$
does not enter the parametrization because it can be expressed
through $\Gamma_\mu(P')$ by the equation analogous to
Eq.~(\ref{Gamma D = D Gamma}):
$$
\frac{1}{\sqrt{s}}\,\Gamma(P)_\mu\,D^j(P,P') =
D^j(P,P')\left\{\frac{1}{\sqrt{s'}}\Gamma_\mu(P') \right.
\left.  -\,\frac{1}{\sqrt{ss'}}\cdot\frac{\sqrt{s'}\,P_\mu +
\sqrt{s}\,P_\mu'} {P_\nu P'\,^\nu +
\sqrt{ss'}}\left[P_\nu\Gamma^\nu(P')\right]\right\}\;.
$$
The set of linearly independent matrices to be used for the
decomposition of the current matrix element is obtained
from the vectors
$P_\mu$ and $\Gamma_\mu(P')$ following Eq.~(\ref{pseud}):
\begin{equation}
D^J(P,\,P')\,(\,P_\mu \Gamma^\mu(P'))^n\;,\quad n = 0,1,\ldots ,2J\;.
\label{pseud2}
\end{equation}

Using the conditions of self--adjointness, current conservation
(\ref{conserv}), parity conservation and orthogonality of the
parametrization vectors to one another (as in
Eq.~(\ref{<|j|>=F_i})) we obtain (see Eq.~(\ref{+}), too):
$$
\langle\vec P,\sqrt s,J,l,S,m_J |j_\mu^{(0)}|\vec P',\sqrt{s'}
,J,l',S',{m'_{J}}\rangle =
$$
\begin{equation}
=  \sum _{m''_J}\,\langle m_J|D^J(P\,,P')|m''_J\,\rangle
\langle\,m''_J|\,\sum ^3_{i=1}\,\left\{F^{ll'SS'}_i\,
A^i_\mu (s,Q^2,s')\right\}_+\,|m'_J\,\rangle\;,
\label{j=FA}
\end{equation}
$$
A^1_\mu =\frac{1}{Q^2}\left[(s-s'+Q^2)P_\mu + (s'-s+Q^2) P\,'_\mu\right]\;,
$$
$$
A^2_\mu =
\frac {1}{\sqrt {s'}}\left\{\,\Gamma_\mu(P') - \frac {1}{2\sqrt s}\left[
-(\sqrt s +\sqrt {s'})\frac {K_\mu}{Q^2}\right.\right.
+ \frac{\sqrt {s'}P_\mu + \sqrt s P'_\mu}{PP' + \sqrt {ss'}} +
$$
$$
\left.\left. +\frac{\sqrt{s}-\sqrt{s'}}{\lambda(s,-Q^2,s')}\left[
(\sqrt{s}+\sqrt{s'})^2 + Q^2\right]\,A^1_\mu\right]
\left[P_\lambda \Gamma^\lambda (P')\right]\right\}\,,
$$
\begin{equation}
A^3_\mu = \frac{i}{\sqrt{s'}}\,R_\mu\;.
\label{Ai}
\end{equation}
The quantities
$F^{ll'SS'}_i$ in Eq.~(\ref{j=FA})
are defined by the relations analogous to
Eq.~(\ref{Fi}):
\begin{equation}
F^{ll'SS'}_i = \sum _{n=0}^{2J}\,f^{ll'SS'}_{in}(s\,,\,Q^2\,,\,s')
(iP_\mu \Gamma^\mu(P'))^n\;.
\label{FilS}
\end{equation}
The sum in Eq.~(\ref{FilS}) is taken using the parity conservation condition as
in Eq.~(\ref{<|j|>=F_i}).

Let us remark that the reduced matrix elements -- the invariant
form factors --- now (in contrast with the form factor
(\ref{<|j|>=A mu g0})) depend on the additional invariant
quantities $l,\,l',\,S,\,S'$ that are invariant
degeneration parameters in the basis
(\ref{bas-cm}).

The self--adjointness condition (\ref{j=FA}) is fulfilled at following
constraints on form factors $f^{ll'SS'}_{in}$:
\begin{equation}
{f^{ll'SS'}}^*_{in}(s, Q^2, s') = f^{l'lS'S}_{in}(s',Q^2, s)\;.
\label{samosop}
\end{equation}
Here the star means the complex conjugation.

Let us consider especially the case
$M_1=M_2=M\,,\,J=J'=S=S'=$1$\,,\,l=l'=$0. The reduced matrix
elements for this case can be used to calculate $\rho$ -- meson
properties neglecting the $D$ -- state contribution. In this case
the functions $F^{ll'SS'}_i$ in Eq.~(\ref{FilS}) have the
following form (compare to Eqs.~(\ref{Fi}), (\ref{<|j|>=F_i})):
$$
F_1 = f_{10}(s,Q^2,s') +
f_{12}(s,Q^2,s')(iP_\nu\Gamma^\nu(P'))^2\;, $$ \begin{equation}
F_2 = f_{21}(s,Q^2,s')\,(i\,P_\nu\Gamma^\nu (P'))\;,\quad
F_3 = f_{30}(s,Q^2,s')\;.
\label{3Fi}
\end{equation}
In equations (\ref{3Fi}) the fixed variables $l,l',S,S'$ are omitted.
Time reflection invariance imposes the additional conditions:
\begin{equation}
f^*_{in} = f_{in}\;,\quad i=1,3\;;
\quad f^*_{21} = -\,f_{21}\;.
\label{otrt}
\end{equation}
With Eqs.~(\ref{samosop}) and (\ref{otrt}) we obtain
\begin{equation}
f_{in}(s, Q^2, s') = f_{in}(s', Q^2, s)\;,
\quad i = 1,3\;;\quad
f_{21}(s, Q^2, s') = -\,f_{21}(s', Q^2, s)\;.
\label{f=f^t}
\end{equation}
The relation (\ref{f=f^t}) demonstrates that the form factor
$f_{21}$ appears in the parametrization as a consequence of the
fact that the invariant masses in the initial and the final
states are different. As will be seen below, this form factor gives no
contribution to elastic processes, for example, to the electron scattering
by the composite particle, however, it does contribute to radiative
transitions.

Let us rewrite Eq.~(\ref{3Fi}) in terms of standard form factors
instead of $f_{in}$:
$$
F_1 = g_{0C}(s,Q^2,s') + g_{0Q}(s,Q^2,s')
\left\{(iP_\nu \Gamma^\nu(P'))^2 \right.
\left. - (1/3)\,
\hbox {Sp}(iP_\nu \Gamma^\nu (P'))^2\right\}
\frac{2}{\hbox {Sp}(P_\nu \Gamma^\nu (P'))^2}\;,
$$
\begin{equation}
F_2 = g_{0MQ}(s,Q^2,s')\,(i\,P_\nu\Gamma^\nu (P'))\;,\quad
F_3 = g_{0M}(s,Q^2,s')\;.
\label{3g}
\end{equation}
The scalar factor in the term with $g_{0Q}$
is chosen in such a way as to make it possible to interpret
$g_{0Q}$ as a quadrupole form factor of the system of two free
particles. In other terms
$g_{0C}$ is the charge form factor, $g_{0MQ}= f_{21}$
is the magnetic quadrupole form factor of the second kind, its
classical analog being the so called toroidal
magnetic moment
\cite{DuT90}, $g_{0M} = f_{30}$ is the magnetic form factor of the free two--
particle system.

In the same way as in the case of
Eq.~(\ref{<|j|>=A mu g0})
the free two--particle form factors in
Eq.~(\ref{3g}) can be written in terms of the constituent form
factors (see Ref. \cite{KrT02} for details). The corresponding equations are
rather complicated, so
we do not present it here.

The free two--particle form factors in Eqs.~(\ref{<|j|>=A mu
g0}), (\ref{j=FA}), (\ref{3Fi}) are to be
considered in the sense of distributions.  For
example, $g_{0}(s\,,Q^2\,,s')$  has
to be interpreted as a Lorentz invariant regular generalized
function on the space of test functions ${\cal S}$({\bf R}$^2$)
\cite{BoL87}.

Let us define the functional giving the regular generalized
function as
\begin{equation}
\langle\,g_{0} (s,Q^2,s')\,,\varphi(s,s')\rangle
= \int\,d\mu(s,s')\,g_{0}(s,Q^2,s')\,\varphi(s,s')\;.
\label{<>}
\end{equation}
Here
\begin{equation}
d\mu(s,s') = 16\,\theta(s - 4M^2)\,\theta(s' - 4M^2)
\sqrt[4]{ss'}\,d\mu(s)\,d\mu(s')\;,\quad
d\mu(s) = \frac{1}{4}k\,d\sqrt{s}\;.\label{dmu}
\end{equation}
The quantity $Q^2$ is a parameter of the generalized function, $M_1=M_2=M$.
The function $\theta(x)$ is the step function.

$\varphi(s\,,s')$ is a function from the space of test functions.
So, for example, the limit as
$Q^2\;\to\;$0 (the static limit) in
$g_0(s\,,Q^2\,,s')$ (\ref{<|j|>=A mu g0})
exists only in the weak sense as the limit of the functional:
\begin{equation}
\lim_{Q^2\to 0}\langle \,g_{0}, \varphi\,\rangle =
\langle(\hbox{e}_q + \hbox{e}_{\bar q})
\delta(\mu(s') - \mu(s)), \varphi\,\rangle\;.\label{lim Q2=0}
\end{equation}
$\hbox{e}_q$ and $\hbox{e}_{\bar q}$ are the constituent
charges, $\delta$ is the Dirac delta--function.

It is just the limit in the sense
(\ref{lim Q2=0}) that gives the electric charge of the free
two--particle system. The ordinary point--wise
limit of the form factor
$g_{0}$ as $Q^2\to$0 is zero.
The equations analogous to Eq.~(\ref{lim Q2=0})
are valid for the static limits of the free two--particle
form factors in
Eqs.~(\ref{3g}), too.

Let us note that the conditions imposed on the free
two--particle form factors that follow from the conditions
for the electromagnetic current operator
(\ref{samosop}), and (\ref{f=f^t})
have to be considered in the weak sense, too.

\section{Parameterization of the current operator matrix
elements for systems of two interacting particles}

In this Section we generalize the parameterization method
of the previous sections to the case of composite system with the
structure defined by the interaction of its
constituents.

Let us consider the operator
$j_\mu(0)$ that describes a transition between two states of a
composite two-- constituent system, $j_\mu(0)$ being diagonal in
the total angular momentum. Let us neglect temporarily the
additional conditions of self--adjointness, parity conservation
etc. in the same way as when constructing the matrix
elements (\ref{<|j|>=F_is}) in Section II.  The Wigner--Eckart
decomposition of the matrix element has the form
(\ref{<|j|>=F_is}), (\ref{Fi}).
To emphasize the fact that the particle is composite, let us
rewrite Eqs.~(\ref{<|j|>=F_is}), (\ref{Fi}) using new notations:
$$
\langle\,\vec p_c,\,m_{Jc}\,|j_\mu(0)|\,\vec p_c\,',m'_{Jc}\,\rangle =
$$
\begin{equation}
= \langle\,m_{Jc}|D^{J_c}(p_c,\,p_c')\,
\left[\,F^c_1\,K'_\mu + F^c_2\,\Gamma_\mu (p_c')  \right.
\left. + F^c_3\,R_\mu + F^c_4\,K_\mu\right]|m'_{Jc}\rangle\;,
\label{<|jc|>=F_is}
\end{equation}
here
\begin{equation}
F^c_i = \sum _{n=0}^{2J_c}\,f^c_{in}(Q^2)(ip_{c\mu}\Gamma^\mu(p_c'))^n\;.
\label{Fic}
\end{equation}
In Eqs.~(\ref{<|jc|>=F_is}), and (\ref{Fic})
$(p_c - p_c')^2 = -\,Q^2\;,\; p_{c\mu}^2=p'_{c\mu}\,^2=M_c^2\;,\; M_c$ is the
mass
of the composite particle. In contrast to Eq. (\ref{<|j|>=F_is}) spins and
masses are omitted in the state vector variables.

In the frame of RHD the form factors of composite systems$f^c_{in}$ are to be expressed in terms of
RHD wave
functions 

In RHD a state of two particle interacting system is described
by a vector in the direct product of two one--particle Hilbert
spaces (see, e.g., Ref.~\cite{KrT02}). So, the matrix element in
RHD can be decomposed in the basis (\ref{bas-cm}):
$$
\langle\vec p_c\,,m_{Jc}|j_\mu(0)|\vec p_c\,'\,,m'_{Jc}\rangle =
$$
$$
\sum\,\int\,\frac{d\vec P\,d\vec P\,'}{N_{CG}\,N_{CG}'}\,
d\sqrt{s}\,d\sqrt{s'}\,\langle\,\vec p_c\,,m_{Jc}|\vec
P\,,\sqrt{s}\,,J\,,l\,,S\,,m_J\rangle\times
$$
$$\times\langle\vec P\,,\sqrt{s}\,,J\,,l\,,S\,,m_J|j_\mu(0)|
\vec P\,'\,,\sqrt{s'}\,,J'\,,l'\,,S'\,,m_{J'}\rangle\times
$$
\begin{equation}
\times\langle
\vec P\,'\,,\sqrt{s'},\,J'\,,l'\,,S'\,,m_{J'}|\vec p_c\,'\,,m'_{Jc}\rangle\;.
\label{j=int}
\end{equation}
Here the sum is over variables $J$,$J'$,$l$,$l'$,$S$,$S'$,$m_J$,$m_{J'}$, and
$\langle\vec P\,'\,,\sqrt{s'}\,J',l',S',m'_J|\vec p_c\,'\,,m'_{Jc}\rangle$ is
the wave function in
the sense of the instant form of RHD.

Let us write the wave function in the form slightly more
general than in Ref.~\cite{KrT02}:
\begin{equation}\langle\vec P\,\,,\sqrt{s}\,,J\,,l\,,S\,,m_{J}|\,\vec p_c\,,m_{J_c}\rangle
= N_C\,\delta (\vec P\, - \vec p_c)\delta_{J_cJ}\delta_{m_{J_c}m_{J}}
\,\varphi^{J_c}_{lS}(k)\;.
\label{wf}
\end{equation}
$$
N_C = \sqrt{2p_{c0}}\sqrt{\frac{N_{CG}}{4\,k}}\;,
$$
The RHD wave function of constituents with
fixed total angular momentum and total spin is defined by the form:
\begin{equation}
\varphi^{J_c}_{lS}(k(s)) =\sqrt{\sqrt{s}(1 - \eta^2/s^2)}\,u_{lS}(k)\,k\;,
\label{phi(s)}
\end{equation}
and is normalized by the condition:
\begin{equation}
\sum_{lS}\int\,u_{lS}^2(k)\,k^2\,dk = 1\;.
\label{norm}
\end{equation}
Here $\eta = M_1^2 - M_2^2\;$,$u_{lS}(k)$  is a model phenomenological wave
function.

To calculate the form factors in
Eqs.~(\ref{<|jc|>=F_is}), and (\ref{Fic})  let us write the
Wigner--Eckart decomposition on the Poincar\'e group for the
current matrix element in the r.h.s. of Eq.~(\ref{j=int}).
However, now there are some difficulties.

The point is that in the previous sections we were dealing with the
parametrization
of local operator matrix elements in the case when the
transformations of the state vectors and of the operators were
defined by  one and the same representation of the quantum
mechanical Poincar\'e group.
While describing the composite systems in RHD a different
situation can arise when the state vectors and the operator
under consideration are transformed following different
representations of this group.

It is just such a situation that takes place in the case of the
matrix element in the r.h.s. of
Eq.~(\ref{j=int}). The operator in the matrix element describes the system of
two
interacting particles and transforms following the
representation with Lorentz boosts generators depending on the
interaction \cite{KrT02}.
The state vectors physically describe the system
of two free particles and present the basis of a representation
with interaction--independent generators. So, the Wigner--Eckart
decomposition in the form used in Section II can not be applied
directly to the matrix element in the integrand in the r.h.s. of
Eq.~(\ref{j=int}). This is caused by the fact that it is
impossible to construct 4--vectors describing the matrix element
transformation properties under the action of Lorentz boosts
from the variables entering the state vectors (contrary to the
case of, e.g., Eq.~(\ref{<|jc|>=F_is})).
In fact, the possibility of matrix
element representation in the form (\ref{<|jc|>=F_is})
is based on the following fact. Let us act by Lorentz
transformation on the operator:
\begin{equation}
\hat U^{-1}(\Lambda)j^\mu(0)\hat U(\Lambda) = \tilde j^\mu(0)\;.
\label{UjU=tj}
\end{equation}
We obtain the following chain of equalities:
$$
\langle \vec p_c\,,m_{Jc}|\tilde j^\mu(0)|\vec p_c\,'\,,m'_{Jc}\rangle
= \langle \vec p_c\,,m_{Jc}|\hat U^{-1}(\Lambda)j^\mu(0)\hat U(\Lambda)
|\vec p_c\,'\,,m'_{Jc}\rangle =
$$
\begin{equation}
= \sum_{\tilde m_{Jc},\tilde m'_{Jc}}
\langle\,m_{Jc}|[D^{J_c}(R_\Lambda)]^{-1}|\,\tilde m_{Jc}\rangle
\langle \Lambda \vec p_c\,,\tilde m_{Jc}|j^\mu(0)|
\Lambda \vec p_c\,'\,,\tilde m'_{Jc}\rangle
\langle\,\tilde m'_{Jc}|D^{J_c}(R_\Lambda)|\,m'_{Jc}\rangle.
\label{tj=LpjLp}
\end{equation}
Here $D^{J_c}(R_\Lambda)$
is rotation matrix realizing the angular momentum
transformation under the action of Lorentz transformations.
The equalities (\ref{tj=LpjLp})
show that the transformation properties of the current as a
4--vector (\ref{UjmuU}) can be described using the 4--vectors
of the initial and the final states. This means that the
procedure of the canonical parameterization can be extended to the matrix
element for the
realization
of the Wigner--Eckart theorem on the Poincar\'e group.

In the case of the current matrix element in the r.h.s. of
Eq.~(\ref{j=int}) the relations
(\ref{tj=LpjLp}) are not valid and direct application
of the Wigner--Eckart theorem is impossible.

However, it can be shown that for the matrix element in
Eq.~(\ref{j=int}) considered as a generalized function (that is
considered as an object having sense only under integrals and
sums in
Eq.~(\ref{j=int})), the equality (\ref{tj=LpjLp}) is valid in the
weak sense.

Let us consider the matrix element in question as a regular
Lorentz covariant generalized function (see, e.g.,
Ref.~\cite{BoL87}). Using Eq.~(\ref{wf}) let us rewrite
Eq.~(\ref{j=int}) in the following form:
$$
\langle\vec p_c\,,m_{Jc}|j_\mu(0)|\vec p_c\,'\,,m'_{Jc}\rangle
= \sum_{l,l',S,S'}\int\,\frac{N_c\,N'_c}{N_{CG}\,N_{CG}'}\,
d\sqrt{s}\,d\sqrt{s'}\,\varphi^{Jc}_{lS}(s)\varphi^{Jc}_{l'S'}(s')\times
$$
\begin{equation}
\times\langle\vec p_c\,,\sqrt{s}\,,J_c\,,l\,,S\,,m_{Jc}|j_\mu(0)|
\vec p_c\,'\,,\sqrt{s'}\,,J_c\,,l'\,,S'\,,m'_{Jc}\rangle\;.
\label{j=int ds}
\end{equation}
Here it is taken into account that the current operator $j_\mu(0)$
is diagonal in total angular momentum of the composite system.

Let us make use of the fact that the set of the states
(\ref{bas-cm}) is complete in RHD:
\begin{equation}
\hat I = \sum\int\,\frac{d\vec P}{N_{CG}}\,d\sqrt{s}\,
|\vec P\,,\sqrt{s}\,, J\,,l\,,S\,,m_{J}\rangle
\langle\vec P\,,\sqrt{s}\,, J\,,l\,,S\,,m_{J}|\;.
\label{I=compl}
\end{equation}
Here the sum is over all the discrete variables of the basis
(\ref{bas-cm}).

Under the integral the matrix element of the transformed current
satisfies the following chain of equalities ((\ref{wf}) and  (\ref{I=compl}) are
taken into
account):
$$
\sum\int\,\frac{N_c\,N_c'}{N_{CG}\,N_{CG}'}\,d\sqrt{s}\,d\sqrt{s'}\,
\varphi^{Jc}_{lS}(s)\varphi^{Jc}_{l'S'}(s')\times
$$
$$
\times
\langle\vec p_c\,,\sqrt{s}\,,J_c\,,l\,,S\,,m_{Jc}|
\hat U^{-1}(\Lambda)j_\mu(0)\hat U(\Lambda)|\vec p_c\,'\,,\sqrt{s'}\,,
J_c\,,l'\,,S'\,,m'_{Jc}\rangle =
$$
$$
=\langle\,\vec p_c\,,m_{Jc}|
\hat U^{-1}(\Lambda)\,\hat I\,j_\mu(0)\,\hat I\,\hat U(\Lambda)
|\,\vec p_c\,'\,,m'_{Jc}\rangle =
$$
$$
=\sum_{\tilde m_{Jc},\tilde m'_{Jc}}
\langle\,m_{Jc}|[D^{J_c}(R_\Lambda)]^{-1}|\,\tilde m_{Jc}\rangle
\langle \Lambda \vec p_c\,,\tilde m_{Jc}|
\hat I\,j^\mu(0)\hat I\,|
\Lambda \vec p_c\,'\,,\tilde m'_{Jc}\rangle
\langle\,\tilde m'_{Jc}|D^{J_c}(R_\Lambda)|\,m'_{Jc}\rangle=
$$
$$
= \sum\int\,\frac{N_c\,N_c'}{N_{CG}\,N_{CG}'}\,
d\sqrt{s}\,d\sqrt{s'}\,\varphi^{Jc}_{lS}(s)\varphi^{Jc}_{l'S'}(s')
\sum_{\tilde m_{Jc},\tilde m'_{Jc}}
\langle\,m_{Jc}|[D^{J_c}(R_\Lambda)]^{-1}|\,\tilde m_{Jc}\rangle\times
$$
\begin{equation}
\times\langle\Lambda\vec p_c\,,\sqrt{s}\,,J_c\,,l\,,S\,,\tilde m_{Jc}|j_\mu(0)
|\Lambda\vec p_c\,'\,,\sqrt{s'}\,,J_c\,,l'\,,S'\,,\tilde m'_{Jc}\rangle
\langle\,\tilde m'_{Jc}|D^{J_c}(R_\Lambda)|\,m'_{Jc}\rangle\;.
\label{j=int3}
\end{equation}

It is easy to see that under the integral the current matrix
element satisfies the equalities analogous to
Eq.~(\ref{tj=LpjLp}), so now it is possible to use the
parameterization method of the previous sections under the
integral, that is to use the Wigner--Eckart theorem in the weak
sense.

The next step is a parameterization of the matrix element in the
r.h.s. of Eq.~(\ref{j=int ds}).  The r.h.s. can be written as a
functional on the space of test functions of the form (see
Eq.~(\ref{phi(s)}, too)):
\begin{equation}
\psi^{ll'SS'}(s\,,\,s') = u_{lS}(k(s))\,u_{l'S'}(k(s'))\;.
\label{psi(ss')}
\end{equation}
Eq.~(\ref{j=int ds}) can be
rewritten as a functional in {\bf R}$^2$ with variables $(s,s')$
(see Eqs.~(\ref{<>})--(\ref{dmu}), too):
$$
\langle\vec p_c\,,m_{Jc}|j_\mu(0)|\vec p_c\,'\,,m'_{Jc}\rangle
= \sum_{l,l',S,S'}\int\,d\mu(s,s')\frac{N_c\,N'_c}{N_{CG}\,N_{CG}'}\,
\psi^{ll'SS'}(s,s')\times
$$
\begin{equation}
\times\langle\vec p_c\,,\sqrt{s}\,,J_c\,,l\,,S\,,m_{Jc}|j_\mu(0)|
\vec p_c\,'\,,\sqrt{s'}\,,J_c\,,l'\,,S'\,,m'_{Jc}\rangle\;.
\label{j=int dmu}
\end{equation}
The measure in the integral (\ref{j=int dmu})
is chosen with the account of the
relativistic density of states, subject to the normalization
(\ref{phi(s)}), (\ref{norm}) (see Eq.~(\ref{dmu}, too)):
$$
d\mu(s,s') = 16\,\theta(s - (M_1+M_2)^2)\,\theta(s' - (M_1+M_2)^2)\,
$$
\begin{equation}
\times\sqrt{\sqrt{s}(1 - \eta^2/s^2)\sqrt{s'}(1 -
\eta^2/{s'\,}^2)}\,d\mu(s)\,d\mu(s')\;.
\label{dmuM1M2}
\end{equation}
$d\mu(s)$ is given by Eq.~(\ref{dmu}).

Notice that the sums over discrete invariant variables can be transformed
into integrals by introducing the adequate delta--functions.
Then the obtained expression is a functional
in {\bf R}$^6$.

The functional in the r.h.s. of
Eq.~(\ref{j=int dmu}) defines a Lorentz covariant generalized
function, generated by the current operator matrix element.
The integral in Eq.~(\ref{j=int dmu})
converges uniformly due to the definition (\ref{psi(ss')}).

Taking into account
Eq.~(\ref{j=int3}) we decompose the matrix element in the r.h.s.
of Eq.~(\ref{j=int dmu}) into the set of linearly independent
scalars entering the r.h.s. of Eq.(\ref{<|jc|>=F_is}):
$$
\frac{N_c\,N'_c}{N_{CG}\,N_{CG}'}\,
\langle\vec p_c\,,\sqrt{s},J_c,l,S,m_{Jc}|j_\mu(0)|\vec
p_c\,'\,,\sqrt{s'},J_c,l',S',m'_{Jc}\rangle =
$$
\begin{equation}
= \langle\,m_{Jc}|D^{J_c}(p_c,\,p_c')
\sum_{n=0}^{2J_c}(ip_{c\mu}\Gamma^\mu(p_c'))^n\,
{\cal A}^{ll'SS'}_{n\mu}(s,Q^2,s')|m'_{Jc}\rangle,
\label{<|jc|>=cA}
\end{equation}
Here ${\cal A}^{ll'SS'}_{n\mu}(s,Q^2,s')$ is a
Lorentz covariant generalized function.

Taking into account
Eq.~(\ref{<|jc|>=cA}) and comparing
the r.h.s. of
Eq.~(\ref{<|jc|>=F_is}) with Eq.~(\ref{j=int dmu}) we obtain:
$$\sum_{l,l',S,S'}\int\,d\mu(s,s')\,
\psi^{ll'SS'}(s,s')
\langle\,m_{Jc}|{\cal A}^{ll'SS'}_{n\mu}(s,Q^2,s')\,|m'_{Jc}\rangle =
$$
\begin{equation}
= \langle\,m_{Jc}|\left[\,f^c_{1n}\,K'_\mu + f^c_{2n}\,\Gamma_\mu (p'_c)
\right.
\left.
+ f^c_{3n}\,R_\mu + f^c_{4n}\,K_\mu\right]|m'_{Jc}\rangle\;.
\label{c=c}
\end{equation}

It is easy to see that all the form factors in the r.h.s. of
Eq.~(\ref{c=c}) are nonzero if the generalized function ${\cal
A}$ contains parts that are diagonal (${\cal A}_1$) and
non-diagonal (${\cal A}_2$) in $m_{Jc}\,,\,m'_{Jc}$:
\begin{equation}
{\cal A}^{ll'SS'}_{n\mu}(s,Q^2,s') = {\cal A}^{ll'SS'}_{1n\mu}(s,Q^2,s') +
{\cal A}^{ll'SS'}_{2n\mu}(s,Q^2,s')\;.
\label{cA=cA1+cA2}
\end{equation}

For the diagonal part we have from
Eq.~(\ref{c=c}):
$$
\sum_{l,l'S,S'}\int\,d\mu(s,s')\,
\psi^{ll'SS'}(s,s')
\langle\,m_{Jc}|{\cal A}^{ll'SS'}_{1n\mu}(s,Q^2,s')\,|m_{Jc}\rangle=
$$
\begin{equation}
= \langle\,m_{Jc}|\left[\,f^c_{1n}[\psi]\,K'_\mu +
f^c_{4n}[\psi]\,K_\mu\right]|m_{Jc}\rangle\;.
\label{cd=cd}
\end{equation}
The notation $f^c_{in}[\psi]$ in the r.h.s. emphasizes the fact
that form factors of composite systems are
functionals on the wave functions of the intrinsic motion
and so, due to Eq.~(\ref{psi(ss')}), on the test functions.

Let the equality
(\ref{cd=cd}) be valid for any test function
$\psi^{ll'SS'}(s,s')$.
When the test functions (the intrinsic motion wave functions)
are changed the vectors in the r.h.s. are not changed because
according to the essence of the parametrization
(\ref{<|jc|>=F_is}) they do not depend on the model for the
particle intrinsic structure. So, when the test functions are
varied the vector of the r.h.s. of Eq.~(\ref{cd=cd})
remains in the hyperplace defined by the vectors
$K_\mu\,,\,K'_\mu$.

When test functions are varied arbitrarily the vector in the
left--hand side (l.h.s.) of Eq.~(\ref{cd=cd}) can take, in
general, an arbitrary direction. So, the requirement of the
validity of Eq.~(\ref{cd=cd}) in the whole space of our test
functions is that the l.h.s. generalized function should have
the form:
\begin{equation}
{\cal A}^{ll'SS'}_{1n\mu}(s,Q^2,s') = K'_\mu\,G^{ll'SS'}_{1n}(s,Q^2,s')
+ K_\mu\,G^{ll'SS'}_{4n}(s,Q^2,s')\;.
\label{cA1}
\end{equation}
Here $G^{ll'SS'}_{in}(s,Q^2,s')\;,\; i=1,4$
are Lorentz invariant generalized functions.

Substituting
Eq.~(\ref{cA1}) in Eq.~(\ref{cd=cd}) and taking into account
Eqs.~(\ref{phi(s)}) and (\ref{dmuM1M2})
we obtain the following integral representations:
\begin{equation}
f^c_{in}(Q^2) =
\sum_{l,l',S,S'}\int_{M_1+M_2}^{\infty}\,d\sqrt{s}\,d\sqrt{s'}\,
\varphi^{Jc}_{lS}(s)\,G^{ll'SS'}_{in}(s,Q^2,s')\varphi^{Jc}_{l'S'}(s')\;.
\label{intrep}
\end{equation}
In the case of matrix element in
Eq.~(\ref{c=c}) non-diagonal in $m_{Jc}\,,\,m'_{Jc}$ we can
proceed in an analogous way and obtain an analogous integral
representations for $f^c_{in}(Q^2)\;,\;i=2,3$.

So, the matrix element in the r.h.s. of
Eq.~(\ref{j=int dmu}) considered as Lorentz covariant
generalized function can be written as the following
decomposition of the type of Wigner--Eckart de\-com\-po\-si\-tion:
$$
\langle\vec p_c\,,\sqrt{s}\,,J_c\,,l\,,S\,,m_{Jc}|j_\mu(0)|
\vec p_c\,'\,,\sqrt{s'}\,,J_c\,,l'\,,S'\,,m'_{Jc}\rangle=
$$
\begin{equation}
= \frac{N_{CG}\,N_{CG}'}{N_c\,N'_c}\,
\langle\,m_{Jc}|D^{J_c}(p_c,\,p_c')\left[\,{\cal F}_{1}\,K'_\mu +
{\cal F}_{2}\,\Gamma_\mu (p'_c) \right.
\left.
+ {\cal F}_{3}\,R_\mu + {\cal F}_{4}\,K_\mu\right]|m'_{Jc}\rangle\;.
\label{fin}
\end{equation}
\begin{equation}
{\cal F}_i = \sum _{n=0}^{2J_c}\,G^{ll'SS'}_{in}(s,Q^2,s')
(ip_{c\mu}\Gamma^\mu(p_c'))^n\;,
\label{cfic}
\end{equation}
with the constraint (\ref{intrep}).

In Eqs.~(\ref{fin}), and (\ref{cfic})
the form factors contain all the information about the physics
of the transition described by the operator
$j_\mu(0)$. They are connected with the composite particle form
factors (\ref{<|jc|>=F_is}), and (\ref{Fic})
through Eq.~(\ref{intrep}). In particular, physical
approximations are formulated in our approach in terms of form
factors $G^{ll'SS'}_{in}(s,Q^2,s')$ (see Ref.~\cite{KrT02} for
details).  The matrix element transformation properties are
given by the 4--vectors in the r.h.s. of Eq.~(\ref{fin}).

It is worth to emphasize that it is necessary to consider the
composite system form factors as the functionals generated
by the Lorentz invariant generalized functions
$G^{ll'SS'}_{in}(s,Q^2,s')$.

After determination of the vectors ${\cal A}^{ll'SS'}_{n\mu}(s,t,s')$ in
Eq. (\ref{c=c}) let us impose the additional conditions
on the matrix
elements in Eqs.~(\ref{<|jc|>=F_is}), and (\ref{c=c})
in the same way as we did in Sec.II in
Eqs.~(\ref{G-K})--(\ref{<|j|>=F_i}). The
r.h.s.  of equalities (\ref{<|jc|>=F_is}) and  (\ref{fin})
contain the same 4--vectors and the same sets of
Lorentz scalars (\ref{Fic}) and (\ref{cfic}), so, to take into
account the additional conditions it is necessary to redefine
these 4--vectors according to
Eqs.~(\ref{G-K})--(\ref{<|j|>=F_i}).  For example, the
conservation law gives ${\cal F}_{4}$=0.  It is easy to see that
for the redefined form factors the equality (\ref{intrep})
remains valid.

Let us write the parameterization
(\ref{fin}), (\ref{cfic}) for the particular case of composite particle
electromagnetic current with quantum numbers
$J=J'=S=S'=1$, which is realized, for example, in the case of
deuteron. Separating the quadrupole form factor in analogy
with Eq.~(\ref{3g}) and using
Eqs.~(\ref{fin}), and (\ref{cfic}) we obtain the following form:
$$
\langle\vec p_c\,,\sqrt{s}\,,J_c\,,l\,,S\,,m_{Jc}|j_\mu(0)|
\vec p_c\,'\,,\sqrt{s'}\,,J_c\,,l'\,,S'\,,m'_{Jc}\rangle =
$$
\begin{equation}
= \frac{N_{CG}\,N_{CG}'}{N_c\,N'_c}\,\langle\,m_{Jc}|\,D^{1}(p_c\,,p'_c)\,
\left[
\tilde{\cal F}_1\,K'_\mu
+ \frac{i}{M_c}\tilde{\cal F}_3\,R_\mu\right]|m'_{Jc}\rangle\;.
\label{J=1}\end{equation}
$$
\tilde{\cal F}_1 = \tilde G^{ll'}_{10}(s,Q^2,s') +
\tilde G^{ll'}_{12}(s,Q^2,s')\left\{[i{p_c}_\nu\,\Gamma^\nu(p'_c)]^2
\right.
\left. - \frac{1}{3}\,\hbox{Sp}[i{p_c}_\nu\,\Gamma^\nu(p'_c)]^2\right\}
\frac{2}{\hbox{Sp}[{p_c}_\nu\,\Gamma^\nu(p'_c)]^2}\;,
$$
\begin{equation}
\tilde{\cal F}_3 = \tilde G^{ll'}_{30}(s,Q^2,s')\;.
\label{FJ=1}
\end{equation}
We have taken into account that the equation
$\tilde G^{ll'}_{21}(s,Q^2,s') =$ 0 is valid in weak
sense.

\section{Conclusion}

The method of construction of the electromagnetic current matrix
elements for the relativistic two--particle composite systems
with nonzero total angular momentum is developed in the frame of
the instant form of RHD.

The method makes use of the Wigner--Eckart theorem on the
Poincar\'e group. It enables one to extract from the matrix
elements the reduced matrix elements -- invariant form
factors -- which in the case of composite systems are
generalized functions.

The obtained current operator matrix elements satisfy the
Lorentz--covariance condition and the conservation law.

The modified impulse approximation (MIA)--- with the physical content
of the relativistic impulse approximation --- was formulated in
Refs.\cite{KrT02,KrT03prc} in
terms of reduced matrix elements. MIA conserves
Lorentz covariance of electromagnetic current and the
current conservation law.

The developed formalism was used in \cite{KrT04} to obtain a reasonable
description of the static moments and the electromagnetic form
factors of $\rho$ meson. A number of relativistic effects are
obtained, for example, the nonzero quadrupole moment (in the
case of $S$ state) due to the relativistic Wigner spin rotation.
In Ref.\cite{KrT03} the tenzor polarization $T_{20} (Q^2)$ in the
elastic $ed$-scattering was calculated. The good agreement with
experiments was obtained, especially for deuteron wave functions from Ref.
\cite{MuT81}. These particular results argue for the validity of our
description of the electroweak properties of composite systems with
nonzero total angular momentum.

This work was supported in part by the Program "Russian
Universities--Basic Researches" \\(Grant No. 02.01.013) and Russian Ministry of
Education (Grant E02--
3.1--34).

\end{document}